# Orbital Kondo effect and spin polarized transport through quantum dots

**S. Lipiński** and **D. Krychowski**

Institute of Molecular Physics, Polish Academy of Sciences, ul. Smoluchowskiego17, 60-179 Poznań, Poland

The coherent spin dependent transport through a set of two capacitively coupled quantum dots placed in a magnetic field is considered within the equation of motion method. The magnetic field breaks the spin degeneracy. For special choices of gate voltages the dot levels are tuned to resonance and the orbital Kondo effect results. For different Zeemann splittings at the dots the Kondo resonance can be formed for only one spin channel. In this case the system operates as an efficient spin filter.

**1  Introduction**    Magnetic nanodevices operating in the coherent regime hold promises for future applications as single-electron transistors or as basic building blocks (qubits) in the fabrication of quantum computers [1]. The advances in nanofabrication techniques opened new path in studying correlation effects. Quantum dots (QDs) offer unprecedented opportunities of realizing and manipulating magnetic impurities interacting with a Fermi sea of conduction electrons. Rich aspects of Kondo physics have been exploited owing to the tunability of relevant parameters in these systems.

In this paper we study double dot (DD) system with strong capacitive interdot coupling [2]. When the two QDs are interacting, the orbital degree of freedom comes into play as a pseudospin, as shown experimentally in Refs. [3-5]. The charge pseudospin can be expressed by the difference of the occupancies of the dots [6]. The Kondo effect has two possible sources in DD: spin and orbital degenecies. When both these degeneracies are simultaneously present the Kondo effect of SU(4) symmetry is realized [6]. Partial breaking of the degeneracy by magnetic field or by difference of gate voltages of the dots results in a crossover to SU(2) Kondo physics either in spin or charge sectors. Application of magnetic field separates the spin channels and we show that for the case when the orbital degeneracy is preserved in one of the channels the system can operate as an efficient spin filter.

**2  Model**    We discuss the system of two capacitively coupled quantum dots placed in a magnetic field. Each of the dots is coupled to the separate pair of the leads. The corresponding Hamiltonian reads:

$$H = \sum_{k\alpha i\sigma} E_{k\alpha i\sigma} c^+_{k\alpha i\sigma} c_{k\alpha i\sigma} + \sum_{i\sigma} E_{i\sigma} d^+_{i\sigma} d_{i\sigma} + U\sum_i n_{i+}n_{i-} + U_1 \sum_{\sigma\sigma'} n_{1\sigma} n_{2\sigma'} + \sum_{k\alpha i\sigma} t_{i\alpha}(c^+_{k\alpha i\sigma} d_{i\sigma} + c.c) \quad (1)$$

where $E_{i\sigma} = E_i + g_i\sigma h$ ( we set $|e| = \mu_B = 1$). The first term describes electrons in the electrodes ($i=1,2$ $\alpha = L,R$), the second represents the field dependent site energies, the third and fourth accounts for intra ($U$) and intercoulomb ($U_1$) interactions and the last one describes the tunneling. In general case, the Zeeman splittings at the dots might be different, what is included by the site dependence of g- factor. The current $I = \Sigma_i I_i = \Sigma_{i\sigma} I_{i\sigma}$ is determined by means of the non-equilibrium Green function method [7] as:

$$I_{i\sigma} = \frac{e}{\hbar} \frac{\Gamma_{iL\sigma}\Gamma_{iR\sigma}}{\Gamma_{iL\sigma} + \Gamma_{iR\sigma}} [n_{iL\sigma} - n_{iR\sigma}] \quad (2)$$

where

$$n_{i\alpha\sigma} = -\frac{1}{\pi} \int_{-D}^{D} d\omega f_\alpha(\omega) \text{Im}[G^r_{i\sigma,i\sigma}] \qquad (3)$$

For simplicity of considerations we restrict here to the case of identical electrodes and to the equal couplings to the dots i.e. $t_{i\alpha} \equiv t$, $\Gamma_{i\alpha\sigma} = \Gamma = \pi t^2 \rho_0$. The formula (2) was derived under the assumption of the elastic transport and the lesser, retarded and advanced bare Green functions in the electrodes are taken as $g^<_{i\alpha\sigma} = 2i\pi f_{i\alpha}$ and $g^{r,a}_{i\alpha\sigma} = \mp i\pi\rho_0$. Here $f_{i\alpha} = f_\alpha$ are Fermi distribution functions of the electrodes and $\rho_0 = 1/2D$ is the assumed constant density of states, $D$ is the half of the bandwidth of electrons in the electrodes. In order to determine the Green function $G^r_{i\sigma,i\sigma}$ we use the equation of motion method (EOM). To truncate the higher order Green functions the self-consistent decoupling procedure proposed by Lacroix is used [8]. This method allows for the description of the system in the whole parameter range. In the limit $(U, U_1) \to \infty$ one gets:

$$G^r_{i\sigma,i\sigma}(\omega) = \frac{1 - \sum_{l\in\Omega}(n_l - H_l(\omega))}{\omega - E_{i\sigma} - \Sigma_0 + \sum_{l\in\Omega}(F_l(\omega) - 2\Sigma_0 H_l(\omega))} \qquad (4)$$

where $\Sigma_0 = -i\Gamma$ is the self-energy for the noninteracting QD due to tunneling of the $i\sigma$ electron, $\Omega = \{(\bar{i}, \sigma), (i, -\sigma), (\bar{i}, -\sigma)\}$, ($\bar{1} = 2, \bar{2} = 1$), $n_l = (n_{lL} + n_{lR})/2$:

$$F_l(\omega) = \Gamma \sum_\alpha \int_{-D}^{D} \frac{d\omega'}{\pi} \frac{f_\alpha(\omega')}{\omega' - \omega - i0^+}, \qquad H_l(\omega) = \Gamma \sum_\alpha \int_{-D}^{D} \frac{d\omega'}{\pi} \frac{f_\alpha(\omega') G^a_{l,l}(\omega')}{\omega' - \omega - i0^+} \qquad (5)$$

Equations (4),(5) and the condition (3) consist a set of self-consistent integral equations, which have to be solved. In the next section we discuss the spin dependent transport of electrostatically coupled quantum dots. To characterize the degree of polarization of conductance (PC) we introduce the following quantity:

$$PC = \frac{C_+ - C_-}{C_+ + C_-} \qquad (6)$$

where $C_\sigma$ denotes the linear conductance $\left.\frac{\partial I_\sigma}{\partial V}\right|_{V\to 0}$.

**3 Numerical results and discussion** We present the results for the Kondo limit $E_1 = -4\Gamma$. $\Gamma$ is taken as the energy unit. The bandwidth of the leads is taken as $D = 50$. In Fig.1 we show the conductance versus magnetic field H and gate voltage difference $\Delta E = E_2 - E_1$ in a gray-scale plot for equal and for different values of g-factors of the dots.

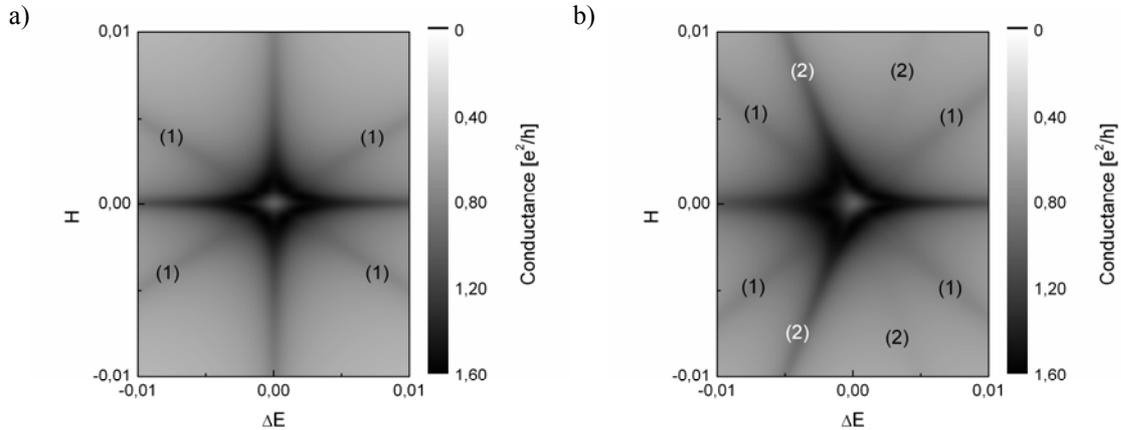

**Fig.1** Conductance vs. magnetic field H and the difference of atomic levels $\Delta E$ for a) $g_1 = g_2 = 1$ b) $g_1 = 1$ $g_2 = 0.5$ calculated for the dot energy $E_1 = -4$.

The high conductance lines (H=0) are clearly visible for both cases. They correspond to the spin Kondo effects at each of the dots ($E_{i+} = E_{i-}$).

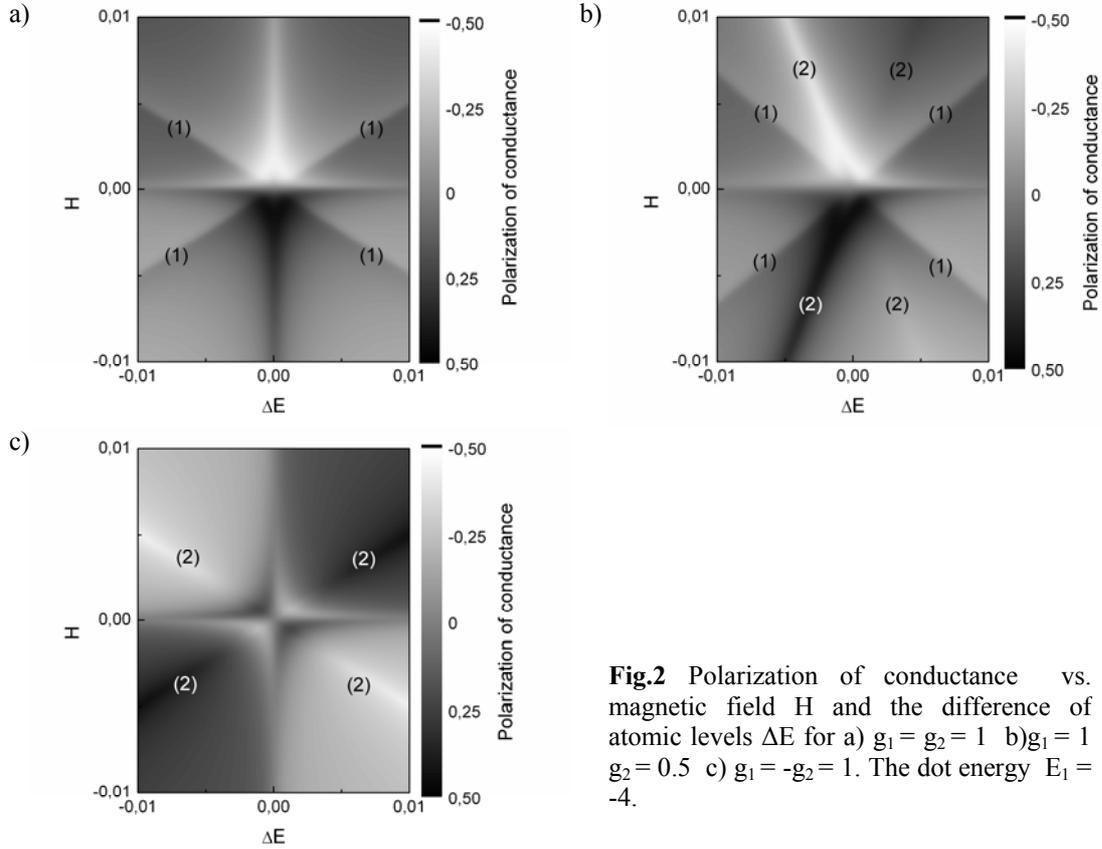

**Fig.2** Polarization of conductance vs. magnetic field H and the difference of atomic levels $\Delta E$ for a) $g_1 = g_2 = 1$  b) $g_1 = 1$ $g_2 = 0.5$  c) $g_1 = -g_2 = 1$. The dot energy $E_1 = -4$.

The high transparency line $\Delta E=0$ is only observed for $g_1 = g_2$ and it reflects the separate orbital Kondo effects for each spin channel (SU(2), $E_{1\sigma} = E_{2\sigma}$). For $\Delta E=0$, H=0 the spin and charge degrees of freedom of the DD are totally entangled (SU(4) Kondo effect). With increasing H ($\Delta E$) Kondo temperature decreases (SU(2)) [6] implying negative magnetoresistance.

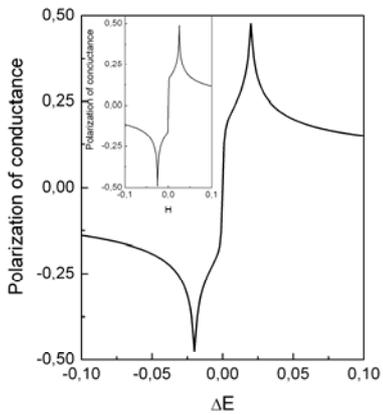

**Fig.3** Polarization of conductance vs. gate voltage $\Delta E = V_{gate}$ for $g_1 = -g_2 = 1$, H = 0.01. Inset shows polarization of conductance vs. magnetic field for $\Delta E = 0.05$.

This effect is reflected in the occurrence of the light spot in the centre diminishing with lowering the position of atomic level. The spin polarized states can be tuned to the orbital degeneracy also for other regions of (H, ΔE) plane. Lines of type (1) represent the range, where the orbital isospin fluctuation processes mix different spin channels (e.g. for ΔE < 0 and H < 0 the $E_{1-} = E_{2+}$ degeneracy results). For $g_1 \neq g_2$ the orbital Kondo effect can also occur in a single spin channel (lines of type 2). The high spin polarization of conductance of lines (2) and much lower polarization of lines 1 is visible on Fig.2. The picture of conductance for the case $g_1 = -g_2$ is identical with the picture for $g_1 = g_2$. The corresponding plots of polarizations of conductance ( Fig2a,c) differ considerably.  The asymmetry of intensities of lines (2) on Figures 1b and 2b for ΔE< 0 and ΔE>0 is a consequence of the fact that for ΔE< 0  the degenerate states of the same spin orientation lie deeper below the Fermi level than for the opposite case. Within the resolution of our gray-scale it is even difficult to identify the lines of type 2 for ΔE>0.  Note that the intensities of lines (1) are symmetric. Figure 3 presents the examples ($g_1 = -g_2$) of individual conductance traces vs. ΔE and vs. H. The system can operate as a bipolar spin filter controlled either by gate voltage or by magnetic field. For the general case of different g factors the situation is similar, but the gate control does not allow for the antisymmetric change of polarization  PC(-ΔE) ≠ -PC(ΔE), what can be still achieved by magnetic field.